\pdfoutput=1

\documentclass[]{raa}           

\usepackage{graphicx,times}
\usepackage{amssymb,amsmath}
\usepackage{color}

\begin{document}

\title{Spectroscopic study of the late B-type eclipsing binary system \\
AR Aurigae A and B: Towards clarifying the differences in 
atmospheric parameters and chemical abundances}

\volnopage{ {\bf 20XX} Vol.\ {\bf X} No. {\bf XX}, 000--000}

\setcounter{page}{1}

\author{Yoichi Takeda}
\institute{11-2 Enomachi, Naka-ku, Hiroshima-shi, 730-0851, Japan; {\it ytakeda@js2.so-net.ne.jp}}

\vs \no
{\small Received 2024 December 10; accepted 2025 January 13}

\abstract{
AR~Aur A+B is a close binary of astrophysical interest, because dissimilar 
surface compositions are reported between similar late B-type dwarfs. 
A new spectroscopic study on this system was carried out based on the 
disentangled spectra, in order to determine their atmospheric parameters 
and elemental abundances.
The effective temperature and microturbulence (determined from the 
equivalent widths of Fe~{\sc ii} lines) turned out (11150~K, 0.9~km~s$^{-1}$) 
and (10650~K, 0.1~km~s$^{-1}$) for A and B.
The chemical abundances of 28 elements were then derived while taking into 
account the non-LTE effect for $Z \le 15$ elements ($Z$: atomic number). 
The following trends were elucidated for [X/H] (abundance of X relative to the Sun): 
(1) Qualitatively, [X/H] shows a rough global tendency of increasing 
with $Z$, with the gradient steeper for A than for B.  
(2) However, considerable dispersion is involved for A, since 
prominently large peculiarities are seen in specific elements reflecting 
the characteristics of HgMn stars (e.g., very deficient N, Al, Sc, Ni; 
markedly overabundant P, Mn).
(3) In contrast, the $Z$-dependence of [X/H] for B tends to be nearly 
linear with only a small dispersion. 
These observational facts may serve as a key to understanding the 
critical condition for the emergence of chemical anomaly. 
}

\authorrunning{Y. Takeda}            
\titlerunning{Atmospheric parameters and abundances of AR~Aur}  
\maketitle

\twocolumn

%
 
\section{Introduction} 

AR~Aur is an eclipsing double-line spectroscopic binary ($P$ = 4.135~d) 
consisting of late B-type stars (B9 + B9.5; hereinafter the former/latter 
primary/secondary components are denoted as A/B, respectively). 
An especially interesting aspect of astrophysical interest regarding this 
system is that the chemical abundances of both components are different 
despite of their apparent similarity. That is, the slightly hotter A 
exhibits a chemically peculiar feature of HgMn type (e.g., detection of 
Hg~{\sc ii} 3984 line) while the other B does not. This distinction in
the surface chemistry between near-identical component stars (which should have
born with the same composition) was recognized already a half century ago 
(Wolff \& Wolff 1976; Takeda et al. 1979; see also sect.~2 in Khokhlova et al. 
1995). Therefore, this binary system can be an important testing bench for 
clarifying the condition for the advent of chemical anomaly. To that end,  
the atmospheric parameters and the elemental abundances for both components 
should be known to a sufficient precision.

Yet, due to the difficulty of measuring individual lines in the spectrum 
(because those of A and B intricately co-exist and merge), only a few 
abundance determinations of AR~Aur A and B have been published so far:
\begin{itemize}
\item
Khokhlova et al. (1995) carefully measured the equivalent widths of A and B
from the double-line spectra in 3850--4600~\AA\ and determined the abundances 
of 18 elements for both components by using the curve-of-growth method. 
\item
Based on the equivalent width data published by Khokhlova et al. (1995),
Ryabchikova (1998) redetermined the abundances of AR Aur A and B (along 
with other HgMn binary stars) by way of the model atmosphere analysis.
\item
Zverko et al. (1997) applied the disentangling technique to obtain the 
individual spectra of A and B in 3910--4630~\AA, and derived the 
abundances of 7 elements for both components by means of spectrum synthesis.
\item
The most extensive and notable work on the chemical abundances of 
AR~Aur A and B is that of Folsom et al. (2010). They employed the 
spectrum-disentangling method to get the separated spectra of A and B 
(4170--6200~\AA), from which the abundances of $\sim 30$
elements were derived by using the spectrum-synthesis method.
\item
Similarly, based on the numerically disentangled spectra of AR~Aur A and B
(along with other four A-type spectroscopic binaries), Takeda et al. (2019;
hereinafter referred to as Paper~I) determined the abundances of 9 elements 
(though the primary attention was paid only to CNO) by applying the synthetic 
spectrum-fitting method.    
\end{itemize}

These previous studies arrived at almost the similar conclusion of distinct 
HgMn-type abundance anomaly in A while a kind of less pronounced 
weak peculiarity in B. But some concern still remains regarding the 
choice of atmospheric parameters.

A potentially important issue is the choice of effective temperature 
($T_{\rm eff}$), which may be one of the critical parameters responsible for 
the onset of chemical anomaly. All of the previous investigations mentioned 
above adopted $T_{\rm eff}$ values of 10950~K (A)\footnote{
In Khokhlova et al.'s (1995) table~4 is given $T_{\rm eff} = 10900$~K 
for AR~Aur~A, although $T_{\rm eff} = 10950$~K is listed in their table~1. 
The reason for this slight difference is not clear.} and 10350~K (B), 
which are the spectrophotometric values (with an uncertainty of $\pm 300$~K) 
determined by Nordstr\"{o}m \& Johansen (1994; see sect.~5 therein) by comparing 
the observed spectral energy distribution (SED) of AR~Aur (3300--7850~\AA) 
with the calculated model flux (A+B combined).
Folsom et al. (2010) independently tried this SED-fitting method 
and obtained $10950 \pm 150$~K (A) and $10350 \pm 150$~K (B); i.e.,
the same results as derived by Nordstr\"{o}m \& Johansen (1994)
even with smaller errors. It may as well be questioned, however,
whether $T_{\rm eff}$'s of two similar stars (especially their difference)
are sufficiently well determinable based on the apparent SED of AR~Aur 
(combined SEDs of both components) alone.   
Therefore, in order to check these results, it would be worthwhile to apply 
the alternative spectroscopic approach, in which $T_{\rm eff}$ is determined 
by requiring the consistency of abundances derived from many lines of 
various lower excitation potential ($\chi_{\rm low}$).    

Another parameter of concern is the microturbulence ($v_{\rm t}$), which 
more or less affects abundance determinations, especially when stronger 
lines are concerned. As a matter of fact, none of the above-mentioned 
past investigations could successfully determine this parameter.
Khokhlova et al. (1995) could estimate only the upper limit of
$v_{\rm t} <1.0$~km~s$^{-1}$ for both A and B, and thus a value 
of 0.5~km~s$^{-1}$ was tentatively assumed in their analysis. 
The same treatment was adopted also by Ryabchikova (1998) and 
Zverko et al. (1997).  Quite similarly, Folsom et al. (2010) could derive 
only the upper limit of 1~km~s$^{-1}$ for A and B, and they eventually 
employed $v_{\rm t} = 0$~km~s$^{-1}$ for both components in their analysis.
In the author's abundance determination in Paper~I, 1.0~km~s$^{-1}$ (A)
and 1.6~km~s$^{-1}$ (B) were tentatively assumed by roughly extrapolating 
the analytical $T_{\rm eff}$-dependent relation empirically derived 
by Takeda et al. (2008) for A-type dwarfs ($7000 \la T_{\rm eff} \la 10000$~K).
Accordingly, given that microturbulence of AR~Aur has never been reliably 
determined  so far, it is desirable to establish this parameter by 
the conventional method using many lines (e.g., Fe lines) by requiring 
that abundances do not show any systematic dependence upon equivalent widths. 

Motivated by this situation, the author decided to conduct a new 
extensive spectroscopic analysis specific to AR~Aur A and B based on 
the disentangled spectra of both components such as done in Paper~I; 
but this time  as much available spectral data as possible are 
exploited in order to make use of as many lines as possible.

Thus, the objectives and intended goals of this study are as follows.
\begin{itemize}
\item
The spectrum-disentangling technique is applied to a set of spectra 
at various phases, in order to obtain disentangled spectra for A and B
covering wide wavelength ranges (from violet to near-IR region) 
used for the analysis.
\item
Then, identification is done for as many spectral lines as possible,
which are of good quality and judged to be usable abundance indicators,  
and their equivalent widths ($W_{\lambda}$) are measured. 
\item
By using these $W_{\lambda}$ of many Fe~{\sc i} and Fe~{\sc ii} lines, 
$T_{\rm eff}$ and $v_{\rm t}$ are spectroscopically determined
by requiring the condition of minimum abundance dispersion (i.e.,
absence of systematic dependence upon these parameters).
\item
Given such established atmospheric parameters, chemical abundances 
of various elements for both A and B are determined based on the 
$W_{\lambda}$ data of identified lines, in order to examine the 
characteristics of surface chemistry and their differences between A and B.  
\item 
In addition, the spectrum-fitting approach is subsidiarily employed 
if necessary (e.g., for the cases of blended lines), and the non-LTE 
effect is taken into account wherever possible.
\end{itemize}

\section{Observational Data}

\subsection{Spectrum disentangling}

As in Paper~I, the basic observational materials employed in this study 
are the high-dispersion spectra of AR~Aur covering wide wavelength ranges 
($\sim$~3800--9200~\AA), which were obtained on 2010 December 14, 15, 
16, 18, and 20 by using BOES (Bohyunsan Observatory Echelle Spectrograph) 
attached to the 1.8 m reflector at Bohyunsan Optical Astronomy Observatory 
(cf. sect.~2.1 and table~2 of Paper~I for more details).

Then, the separated spectra of A and B were numerically obtained by applying 
the spectrum-disentangling technique to this set of original double-line (A+B) 
spectra at 11 different phases.
For this purpose, the public-domain program CRES\footnote{http://sail.zpf.fer.hr/cres/} 
written by Dr. S. Iliji\'{c} was employed with the same procedure as 
detailed in sect.~2.2 of Paper~I. 
Yet, since this study intended to
make use of spectra of wide wavelength region as available as possible,
in contrast to the case of Paper~I  (where only selected 4 narrow regions 
were relevant),  some specific considerations had to be made in  
preconditioning the original spectra to be disentangled.
\begin{itemize}
\item
The existence of very strong lines with broad wings (such as H lines of 
Balmer or Paschen series) makes the disentangling process complicated. In 
order to circumvent this problem, the continuum normalization of all 
original spectra was done by regarding these wings of strong lines
as if being the pseudo-continuum level, so that these features may be 
conveniently wiped out in the spectra. However, since this procedure  
does not work well in/near the sharply changing core regions of such lines, 
they had to be abandoned.     
\item
Another problem of nuisance is the existence of telluric lines, which 
are appreciably seen in specific spectral regions (especially in longer 
wavelength ranges). These telluric features in the original spectra were 
either removed in advance by dividing them by the spectrum of a rapid 
rotator or erased interactively by hand on the screen (if they are weak 
and not so many). However, those regions being dominated by very 
strong and numerous telluric lines had to be discarded, since their 
adequate removal turned out practically impossible.
\end{itemize}

In consequence, disentangled spectra of AR~Aur A and B in 69 spectral regions 
(partially overlapped with each other) were obtained, which cover from 
$\sim 3900$~\AA\ to $\sim 9200$~\AA\ with a step of $\sim$~0.05--0.1~\AA.
The resulting spectra are graphically displayed in Figure~1a and 1b,
where strong H lines are not seen any more and narrow spectral gaps 
(abandoned regions) are observed here and there. Likewise, how the 
signal-to-noise ratio (typically $\sim$~200--600) depend upon the wavelength 
is also shown in Figure~1c and 1d, where we can see that a maximum is 
attained around $\sim 6000$~\AA.    
All these spectra used for the analysis are included in the online 
materials (``spectra\_A.txt'' and ``spectra\_B.txt''). 

\begin{figure}[h] 
\begin{minipage}{70mm}
\begin{center}
   \includegraphics[width=7.0cm]{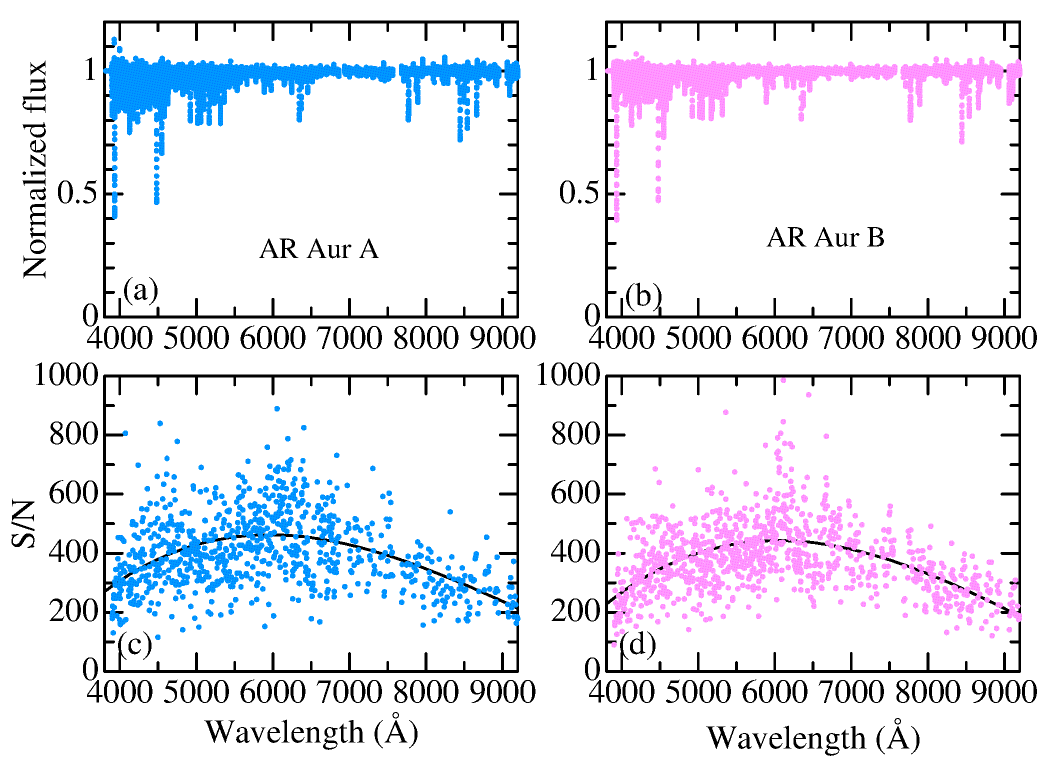}
\end{center}
\caption{The whole disentangled spectra of AR~Aur A and B used in this 
study are plotted against wavelength in the upper panels (a) and (b).
The runs of signal-to-noise ratios of these spectra (directly estimated 
in the line-free regions) are shown in the lower panels (c) and (d), 
where the fitted 3rd-order polynomials (describing the global trends) 
are also depicted by solid lines. 
} 
   \label{Fig1}
\end{minipage}
\end{figure}

\subsection{Line identification and equivalent widths}

Based on the disentangled spectra, lines usable for abundance determinations
were identified and their equivalent widths were measured.
The identification was done by carefully comparing the observed spectrum with 
the theoretically synthesized one. Here, lines to be measured were restricted 
to only those of single component (i.e., multiplet lines with fine structure 
such as Mg~{\sc ii} 4481 were discarded), and those being seriously affected 
by blending with other neighborhood lines were avoided.

Since spectral line shapes of AR~Aur are rather rounded (reflecting the 
projected rotational velocity of $v_{\rm i}\sin i \simeq 23$~km~s$^{-1}$; 
cf. table~1 of Paper~I), the equivalent width ($W_{\lambda}$) of each line 
was evaluated by fitting its profile with a specifically devised function, 
which was constructed by convolving the rotational broadening function 
with the Gaussian function in an appropriately adjusted proportion.
Figure~2 displays the actual examples of how the equivalent widths were 
measured by function-fitting for selected 24 lines of different elements.
All the data ($W_{\lambda}$ along with the atomic data taken from VALD database;
cf. Ryabchikova et al. 2015) of finally identified 606/538 lines for A/B 
are also presented as the online material (``identlist\_A.dat'' and 
``identlist\_B.dat''). 

\begin{figure}[h]
\begin{minipage}{70mm}
\begin{center}
  \includegraphics[width=7.0cm]{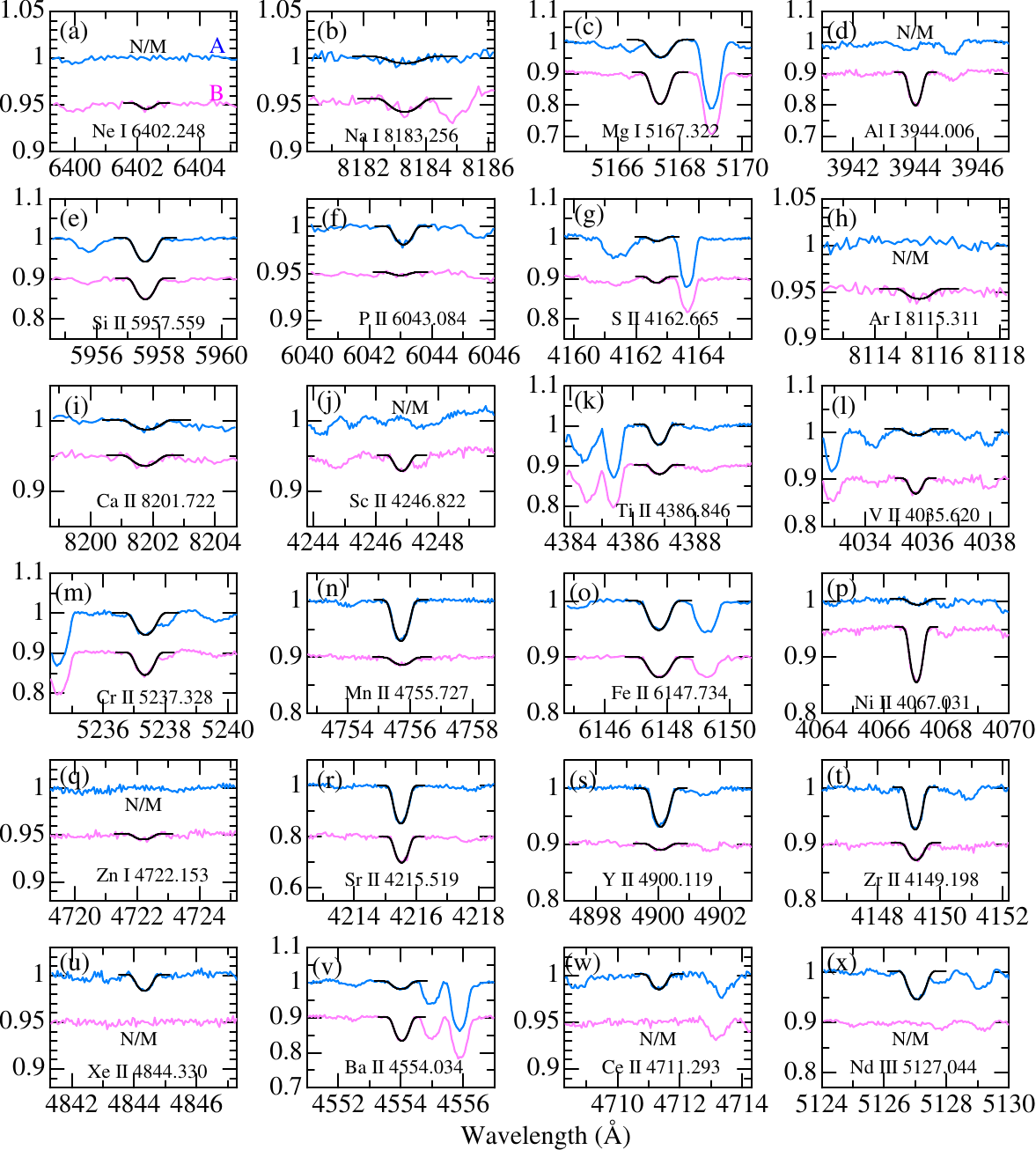}
\end{center}
   \caption{
In order to demonstrate how the equivalent widths were measured by applying 
the direct function-fitting method to those spectral lines identified and 
judged to be usable, selected cases of 24 lines (as indicated in each panel) 
are shown here. The observed spectra are shown in color (blue for A, pink for B), 
while the fitted function is depicted by black solid lines. Note that the scale 
marked in the ordinate is for A, since the spectra for B are shifted downwards by 
appropriate amounts (0.05 or 0.1 or 0.2). The cases where the line is too weak 
to be detected are indicated as ``N/M''(Not Measurable).
} 
   \label{Fig2}
\end{minipage}
\end{figure}

\section{Atmospheric Parameters}

\subsection{Determination method}

As mentioned in Section~1, one of the main objectives of this study 
is to establish $T_{\rm eff}$ and $v_{\rm t}$ for both AR~Aur A and B 
by making use of many Fe lines (which are most numerously available).
According to the traditional approach, $T_{\rm eff}$ can be determined by 
the requirement that the abundances ($A_{i}$) derived from the equivalent 
widths ($W_{i}$) of each line $i$ ($i = 1, 2, \ldots, N$; where $N$ is 
the number of lines) do not systematically depend upon the lower 
excitation potential ($\chi_{\rm low}$) [excitation equilibrium], 
while $v_{\rm t}$ is determinable by demanding that $A_{i}$'s do not 
show any systematic trend irrespective of the strengths of lines 
($W_{i}$) [curve-of-growth matching].
The desired solution ($T_{\rm eff}^{*}, v_{\rm t}^{*}$) simultaneously 
satisfying these two conditions can be obtained by finding the minimum 
of $\sigma(T_{\rm eff}, v_{\rm t})$, where $\sigma$ is the standard 
deviation of the abundances around the mean ($\langle A \rangle$) 
calculated for various combinations of $(T_{\rm eff}, v_{\rm t})$.  

The numbers of Fe~{\sc i} and Fe~{\sc ii} lines ($N_{1}$, $N_{2}$) among 
the list of identified lines (cf. Sect.~2.2) are (64, 306) for A and
(90, 245) for B.
For each species and each star, Fe abundances ($A_{i}, i=1, 2, \cdots, N$) 
were calculated from the equivalent widths ($W_{i}, i=1, 2, \cdots, N$) 
for an extensive grid of 1071 (=51$\times$21) 
cases resulting from combinations of 51 $T_{\rm eff}$ (from 9500 to 12000~K 
with a step of 50~K) and 21 $v_{\rm t}$ (from 0.0 to 2.0~km~s$^{-1}$ with 
a step of 0.1~km~$^{-1}$), where the necessary model atmospheres 
(solar metallicity models with $\log g = 4.3$)\footnote{Throughout this study,
$\log g= 4.3$ is adopted for the surface gravity of both A and B, and solar 
metallicity models are used (cf. Sect.~3.2).} were generated by interpolating 
Kurucz's (1993) grid of ATLAS9 models. Then, $\langle A \rangle$ and $\sigma$
are calculated from the resulting set of ($A_{i}, i=1, 2, \cdots, N$)
for each of the 1071 combinations of ($T_{\rm eff}, v_{\rm t}$),
while outlier data (judged by Chauvenet's criterion) were discarded. 

The contour maps of such obtained $\sigma_{1{\rm A}}$, $\sigma_{2{\rm A}}$,
$\sigma_{1{\rm B}}$, and $\sigma_{2{\rm B}}$ on the $T_{\rm eff}$--$v_{\rm t}$
plane are depicted in Figure~3a, b, c, and d, respectively.
The solutions of ($T_{\rm eff}^{*}$, $v_{\rm t}^{*}$) at the minimum of 
$\sigma$ (denoted by crosses in Fig.~3) for each case are 
summarized in Table~1.

The resulting ($T_{\rm eff}^{*}/v_{\rm t}^{*}/\langle A \rangle$) 
derived from Fe~{\sc i} and Fe~{\sc ii} lines are read from Table~1  as
(10950/1.2/7.89)$_{1{\rm A}}$ and (11150/0.9/8.09)$_{2{\rm A}}$ for A, while 
(9900/0.0/7.30)$_{1{\rm B}}$ and (10650/0.1/7.79)$_{2{\rm B}}$ for B,
where the values are rounded in consideration of the uncertainties.  
Therefore, while two kinds of results from Fe~{\sc i} and Fe~{\sc ii} 
lines are not much different from each other for A, marked discordance 
is observed for the case of B in the sense that $T_{\rm eff}$ and 
$\langle A \rangle$ from Fe~{\sc i} (9900~K, 7.30) are considerably 
lower than that from Fe~{\sc ii} (10650~K, 7.79).

\begin{figure}[h]
\begin{minipage}{70mm}
\begin{center}
  \includegraphics[width=7.0cm]{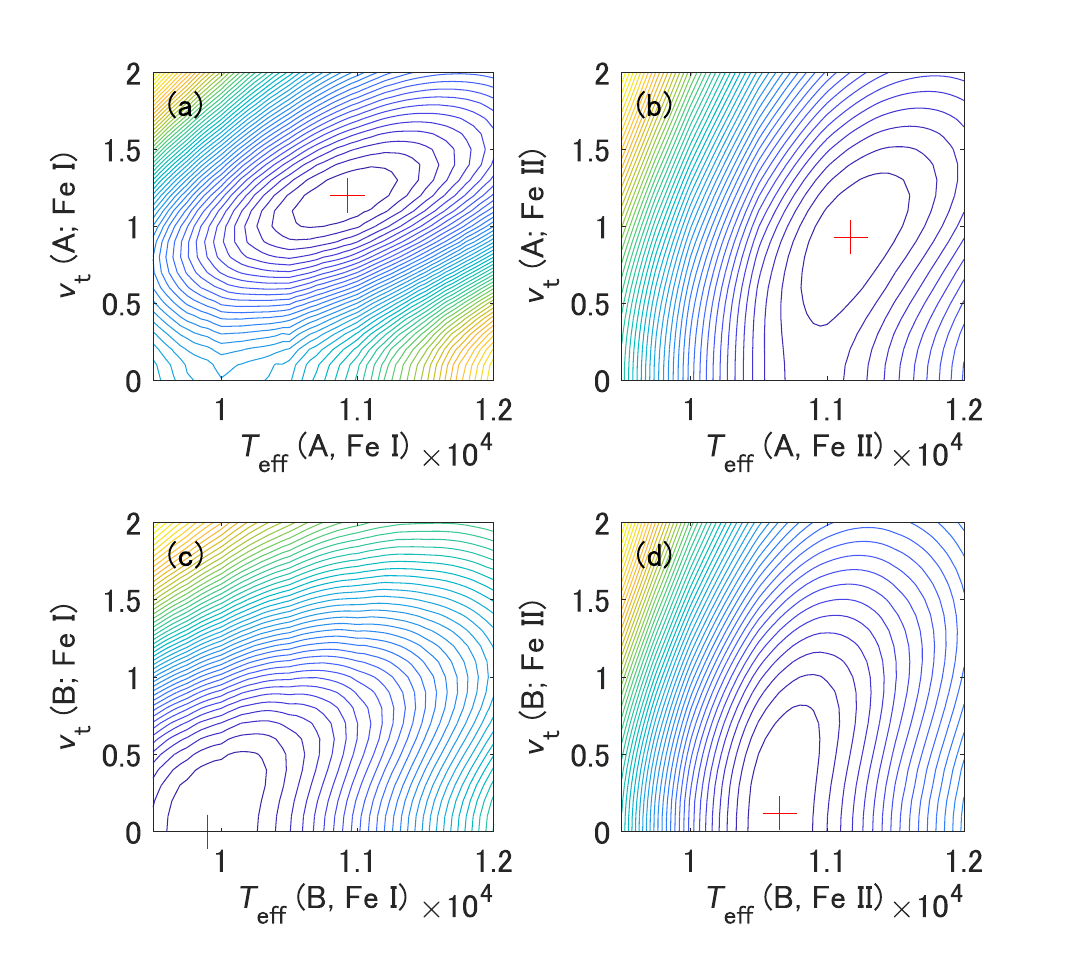}
\end{center}
   \caption{
Graphical display of the contours of $\sigma(T_{\rm eff}, v_{\rm t})$, 
where the position of $(T_{\rm eff}^{\rm min}, v_{\rm t}^{\rm min})$ 
corresponding the minimum of $\sigma$ is indicated by a cross.
The upper panels (a, b) are for AR~Aur A, while the lower ones
(c, d) are for AR~Aur B. The left panels (a, c) and the right ones (b, d)
are for Fe~{\sc i} and Fe~{\sc ii}, respectively. 
} 
   \label{Fig3}
\end{minipage}
\end{figure}

\subsection{Adopted parameters}

Although two kinds of ($T_{\rm eff}^{*}, v_{\rm t}^{*}$) solutions were 
derived in Section~3.1 from Fe~{\sc i} and Fe~{\sc ii} lines, the latter
Fe~{\sc ii} results are considered to be more reliable than the former 
Fe~{\sc i} ones for several reasons:
(i) Since only an extremely tiny fraction of Fe atoms remain neutral 
in the atmosphere of late B-type stars, the formation of Fe~{\sc i} lines 
is considerably $T$-dependent and parameter solutions from them are 
vulnerable to slight inadequacies in observational data, while Fe~{\sc ii} 
lines are more robust in this respect. (ii) The number of Fe~{\sc ii} lines
is by $\sim$~3--4 times as that of Fe~{\sc i} lines, thus the former being
expected to yield statistically more reliable results.
(iii)  A much wider span of $\chi_{\rm low}$ is covered by Fe~{\sc ii} lines
($0 \la \chi_{\rm low} \la 13$~eV) than the case of Fe~{\sc i} lines
($0 \la \chi_{\rm low} \la 5$~eV), which means that the former is definitely 
more advantageous than the latter for the purpose of $T_{\rm eff}$ determination. 
Accordingly, the ($T_{\rm eff}^{*}, v_{\rm t}^{*}$) results derived from 
Fe~{\sc ii} lines are adopted; i.e., (11150, 0.9) for A, and (10650, 0.1) for B.

Regarding the surface gravity, $\log g = 4.30$ is assumed for both A and B 
in this study. Although Nordstr\"{o}m \& Johansen (1994) derived
$4.331(\pm 0.025)$ (A) and $4.280(\pm 0.025)$, their conclusion of 
$\log g_{\rm A} > \log g_{\rm B}$ (which is significant in 
understanding the evolutionary status of this system) appears questionable 
as discussed in Appendix~A. Accordingly, $\log g = 4.30$ was  
assigned to both A and B with an uncertainty of $\la 0.05$~dex,
which is practically sufficient because abundances are less
sensitive to this parameter.
  
As to the model metallicity, solar composition models are adopted,
which is a reasonable choice (given that abundances of some elements are 
subsolar while those of others are supersolar in chemically peculiar stars) 
because atmospheric structures of early-type stars do not depend much 
upon the metallicity (i.e., electrons are donored not by metals but 
essentially by hydrogen).  

The final parameters of AR~Aur A and B used 
for abundance determinations are summarized in Table~2, where 
probable uncertainties are also given.  
The Fe abundances ($A_{i}$) derived from each of Fe~{\sc i} and 
Fe~{\sc ii} lines corresponding to the adopted Fe~{\sc ii}-based 
$T_{\rm eff}^{*}/v_{\rm t}^{*}$ (11150/0.9 for A and 10650/0.1 for B) 
are plotted against ($W_{i}$) and $\chi_{\rm low}$ and the corresponding 
empirical curves of growth are depicted in the center (Fe~{\sc i}) and 
right (Fe~{\sc ii}) columns of Figure~4 (A) and Figure~5 (B),
where the Fe~{\sc i} results for the unadopted Fe~{\sc i} lines-based 
parameters are also shown in the left columns for reference.
It can be seen from these figures that the required condition 
(no systematic dependence in $A_{i}$ upon $W_{i}$ and $\chi_{\rm low}$)
is almost fulfilled.

\setcounter{table}{0}
\begin{table}[h]
\begin{minipage}{70mm}
\begin{center}
\caption{($T_{\rm eff}$, $v_{\rm t}$) solutions at the minimum of $\sigma$.}
\small
\begin{tabular}{cccccc}
\hline\hline\noalign{\smallskip}
Lines & $T_{\rm eff}^{*}$ & $v_{\rm t}^{*}$  &  
$\sigma^{*}$ & $\langle A \rangle$ & Fig. \\
      &  (K)          & (km~s$^{-1}$)  &  (dex)     & (dex)   &  \\
\hline\noalign{\smallskip}
\multicolumn{6}{c}{[AR Aur A]} \\
 Fe~{\sc i}    & 10927  &   1.20  &   0.102 &  7.892 & 3a \\ 
               & (397)  &  (0.23) &         &        &    \\
 Fe~{\sc ii}   & 11172  &   0.93  &   0.143 &  8.090 & 3b \\
               & (187)  &  (0.23) &         &        &    \\
\hline\noalign{\smallskip}
\multicolumn{6}{c}{[AR Aur B]} \\
 Fe~{\sc i}    & 9901  &    0.00  &   0.085 &  7.301 & 3c \\ 
               & (283)  &  (0.17) &         &        &    \\
 Fe~{\sc ii}   & 10655  &   0.12  &   0.143 &  7.790 & 3d \\
               & (140)  &  (0.19) &         &        &    \\
\hline
\end{tabular}
\end{center}
Columns 2 and 3 give the values of $T_{\rm eff}$ and $v_{\rm t}$,  
at which Fe abundance dispersion is minimized. The corresponding 
$\sigma$ and the mean Fe abundance are presented in columns 4
and 5, respectively.  See the figures indicated in column 6 
for the relevant $\sigma (T_{\rm eff}, v_{\rm t})$ contours.
The parenthesized values are the probable uncertainties involved
in $T_{\rm eff}^{*}$ and $v_{\rm t}^{*}$, which were estimated from 
$\sigma^{*}$ by random simulations as described in sect.~3.3 of 
Takeda (2024a). 
\end{minipage}
\end{table}

\setcounter{table}{1}
\begin{table}[h]
\begin{minipage}{70mm}
\begin{center}
\caption{Adopted atmospheric parameters.}
\small
\begin{tabular}{cccc}
\hline\hline\noalign{\smallskip}
Star & $T_{\rm eff}$ & $v_{\rm t}$ & $\log g^{\#}$ \\
     &  (K)          & (km~s$^{-1}$ & (dex)   \\
\hline\noalign{\smallskip}
A    & 11150        &        0.9   &    4.30      \\ 
     & ($\pm 200$)  &  ($\pm 0.2$) & ($\pm 0.05$) \\
\hline\noalign{\smallskip}
B    & 10650        &        0.1   &    4.30      \\
     & ($\pm 150$)  &  ($\pm 0.2$) & ($\pm 0.05$) \\
\hline
\end{tabular}
\end{center}
Given here are the model atmosphere parameters (based on the solutions 
obtained from Fe~{\sc ii} lines; cf. Table~1) finally adopted for 
deriving the chemical abundances of AR Aur A and B.
The parenthesized values are the estimated typical 
uncertainties.\\
$^{\#}g$ is in cm~s$^{-2}$.
\end{minipage}
\end{table}

\begin{figure}[h] 
\begin{minipage}{70mm}
\begin{center}
   \includegraphics[width=7.0cm]{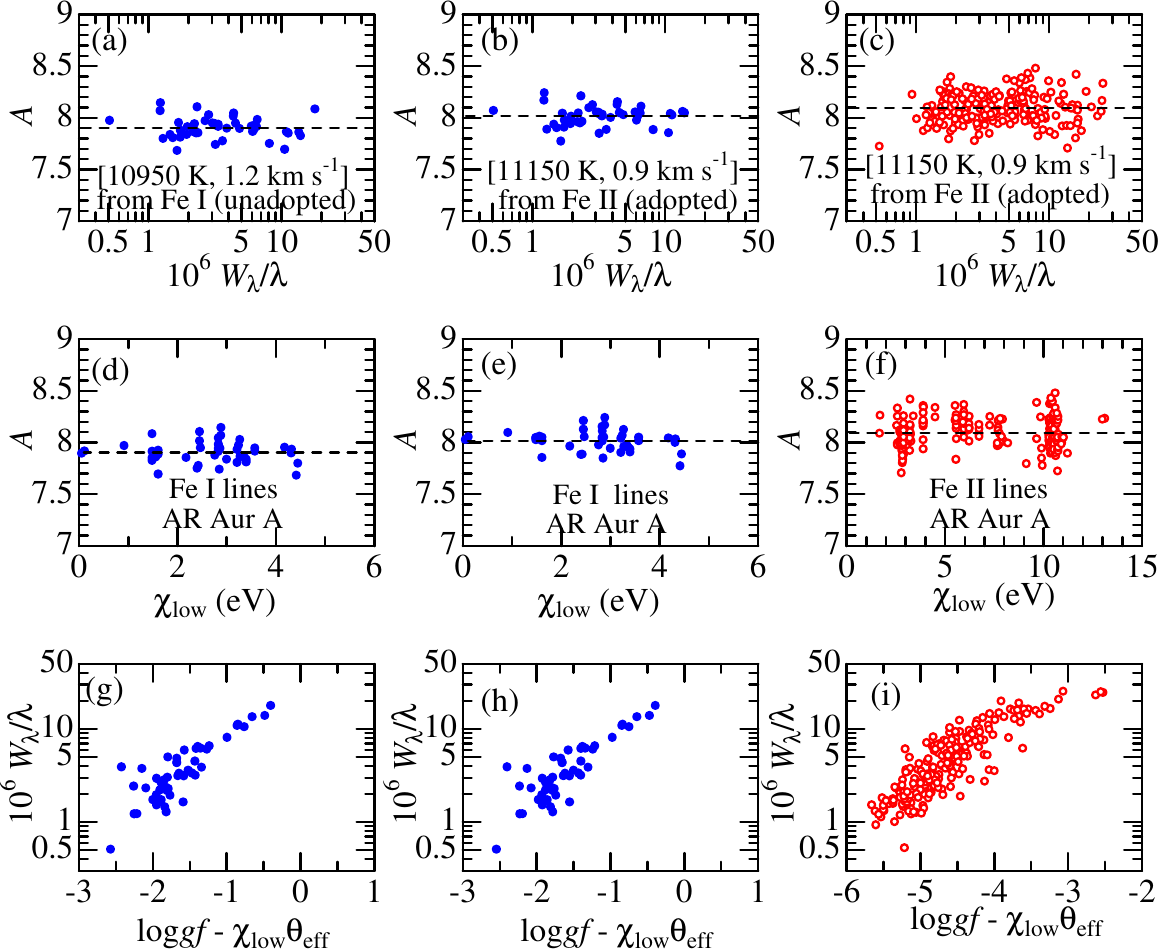}
\end{center}
\caption{
Fe abundances ($A_{i}$) of AR Aur~A corresponding to two ($T_{\rm eff}$, $v_{\rm t}$) 
solutions of minimizing $\sigma$, (10950~K, 1.2~km~s$^{-1}$ based on Fe~{\sc i} 
lines though unadopted; left panels) and (11150~K, 0.9~km~s$^{-1}$ based on Fe~{\sc ii} 
lines finally adopted; center and right panels), are plotted against equivalent width 
($W_{\lambda}$; top panels) or lower excitation potential ($\chi_{\rm low}$;
middle panels). The mean abundance ($\langle A \rangle$) is also indicated by the 
horizontal dashed line. In the bottom panels are shown the corresponding empirical 
curves of growths, where $\log gf - \chi_{\rm low}(5040/T_{\rm eff})$ is taken 
as the abscissa. The results for Fe~{\sc i} and Fe~{\sc ii} lines are distinguished
by filled blue symbols and open red symbols, respectively.
} 
   \label{Fig4}
\end{minipage}
\end{figure}

\begin{figure}[h] 
\begin{minipage}{70mm}
\begin{center}
   \includegraphics[width=7.0cm]{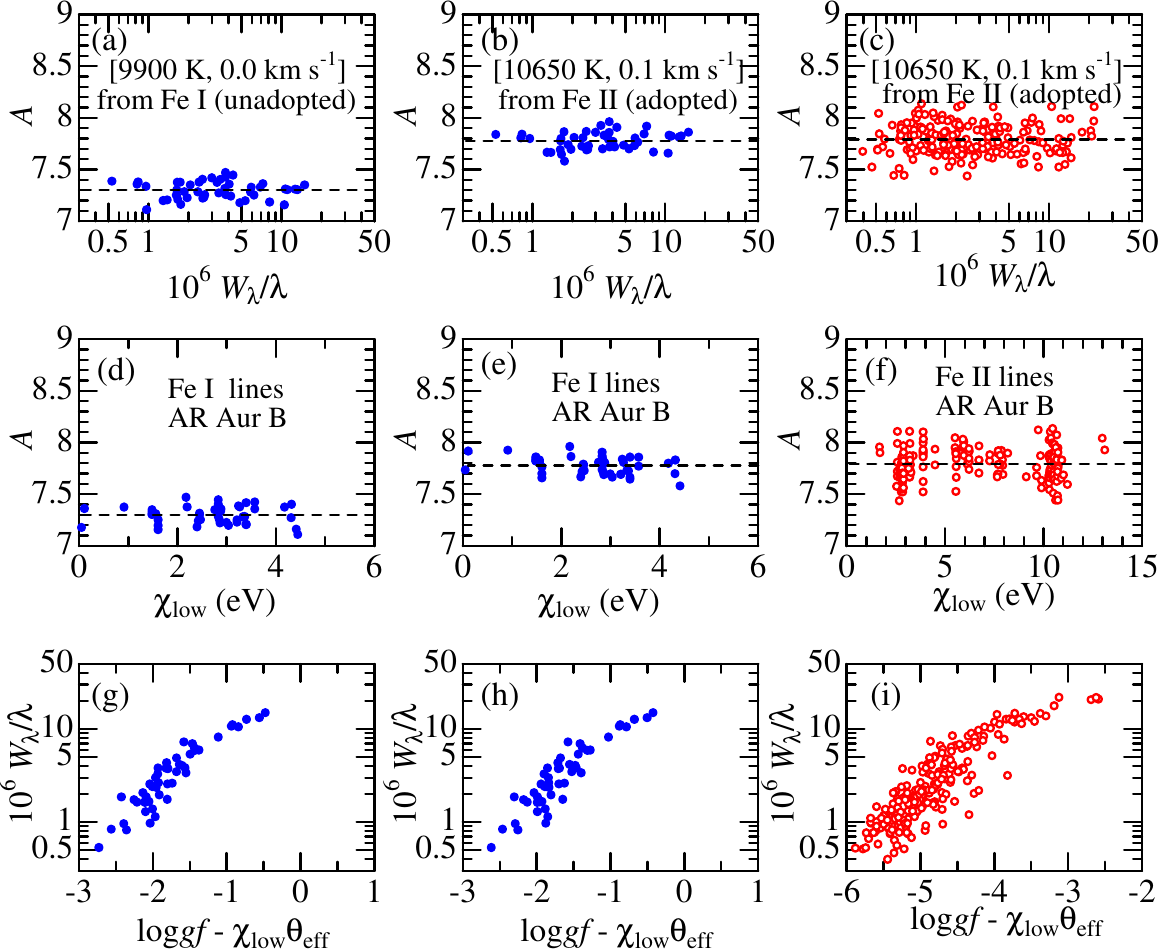}
\end{center}
\caption{
The results of Fe abundances ($A_{i}$) for AR Aur~B are presented, corresponding 
to two ($T_{\rm eff}$, $v_{\rm t}$) solutions of minimizing $\sigma$, 
(9900~K, 0.0~km~s$^{-1}$ based on Fe~{\sc i} lines though unadopted; left panels) 
and (10650~K, 0.1~km~s$^{-1}$ based on Fe~{\sc ii} lines finally adopted; 
center and right panels). Otherwise, the same as in Figure~4.
} 
   \label{Fig5}
\end{minipage}
\end{figure}

\section{Abundance determination}

Based on the model atmospheres with the atmospheric parameters 
established in Section~3.2, elemental abundances of A and B are 
determined from the equivalent widths ($W_{\lambda}$) of spectral lines,
where $W_{\lambda}$ values measured by direct function-fitting on the 
spectrum (cf. Sect.~2.2) are basically employed. 

\subsection{Spectrum fitting}

However, since identification and $W_{\lambda}$ measurement done in 
Section~2.2 was restricted to single-component lines. important 
multi-component line features (such as He~{\sc i} 4471 or Mg~{\sc ii} 4481) 
could not be included. 

Therefore, an alternative synthetic spectrum-fitting approach 
was additionally applied to selected line features (consisting of 
single- or multi-components) to evaluate the relevant $W_{\lambda}$
inversely from the resulting abundance.
Regarding the details of this alternative $W_{\lambda}$-determination
approach, section~4 of Takeda et al. (2018) may be consulted. 

This synthetic fitting analysis was applied to 12 spectral regions,
in order to evaluate the equivalent widths of 18 line features,
as graphically displayed in Figure~6.

\begin{figure}[h]
\begin{minipage}{70mm}
\begin{center}
  \includegraphics[width=7.0cm]{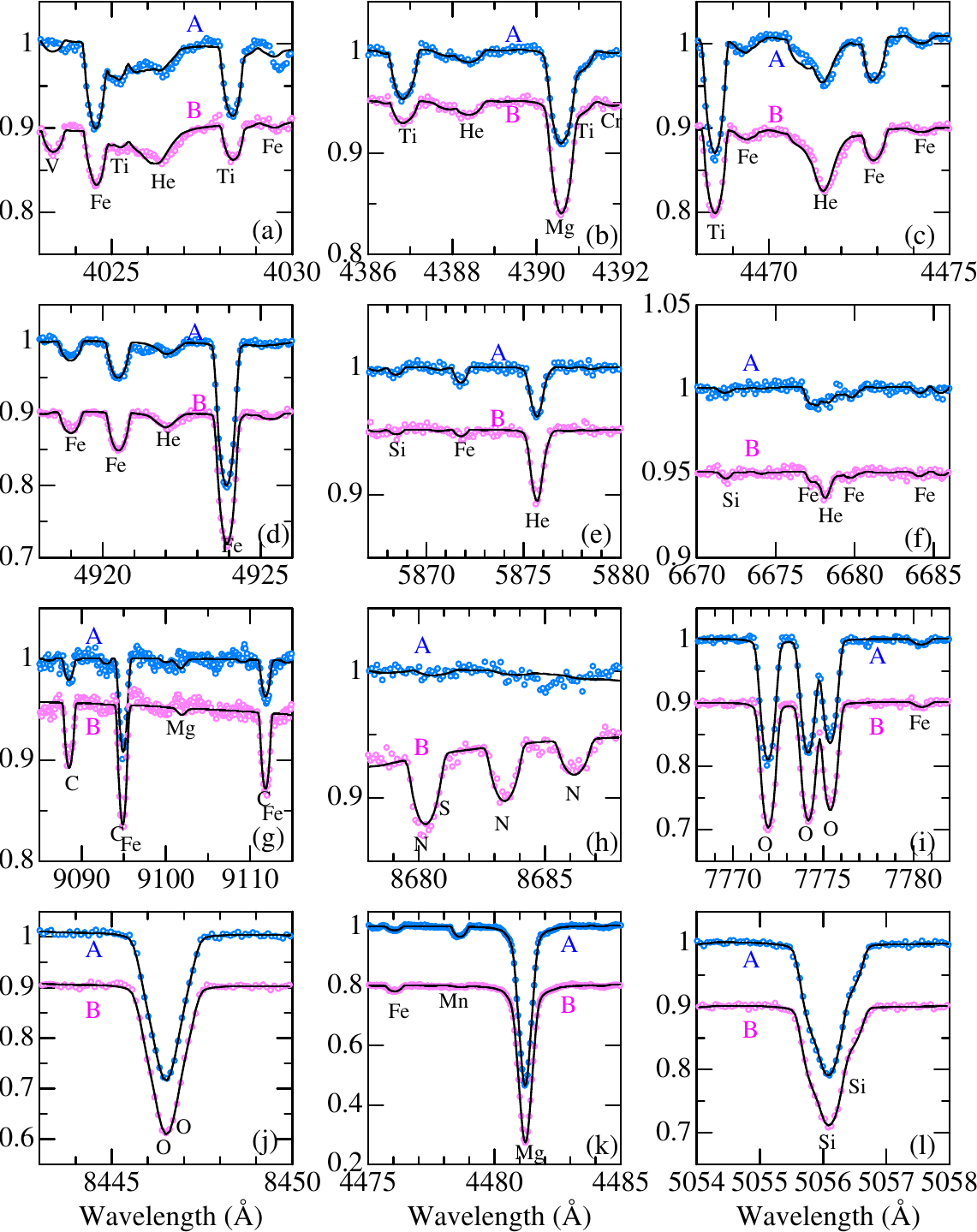}
\end{center}
   \caption{
The accomplished fit of the synthetic spectrum-fitting analysis 
carried out in 12 regions for the purpose of inverse evaluations
of equivalent widths for (a) He~{\sc i} 4026, (b) He~{\sc i} 4388,
(c) He~{\sc i} 4471, (d) He~{\sc i} 4922, (e) He~{\sc i} 5876,
(f) He~{\sc i} 6678, (g) C~{\sc i} 9089/9095/9112, 
(h) N~{\sc i} 8680/8683/8686, (i) O~{\sc i} 7772/7774/7775,
(j) O~{\sc i} 8446, (k) Mg~{\sc ii} 4481, and (l) Si~{\sc ii} 5056.
The observed and theoretical spectra are depicted by symbols 
(blue for A, pink for B) and black lines, respectively. 
As in Figure~2,  the scale marked in the ordinate is for A, since 
the spectra for B are shifted downwards by appropriate amounts 
(0.05 or 0.1 or 0.2). 
} 
   \label{Fig6}
\end{minipage}
\end{figure}

\subsection{Non-LTE calculations}

According to the policy of taking the non-LTE effect into consideration 
wherever possible, non-LTE abundances were derived for comparatively lighter 
elements ($Z \le 15$), for which the author has experiences of non-LTE analysis. 
Otherwise ($Z \ge 16$), the abundances were derived under the assumption of LTE. 
The finally adopted (input) abundances in non-LTE calculations (and the related 
papers) are summarized in Table~3.

Since magnesium (Mg, $Z = 12$) has not been explicitly mentioned in 
the author's past publications, some explanations are in order. 
Non-LTE calculations for Mg~{\sc i} and Mg~{\sc ii} were conducted 
for this study by using the non-LTE code described in Takeda (1991). 
The atomic model was constructed based on Kurucz \& Bell's (1995) 
compilation of atomic data ($gf$ values, levels, etc.), which consists of 
108 Mg~{\sc i} terms (up to 19$d$~$^{3}$D at 61361.5~cm$^{-1}$) 
with 771 Mg~{\sc i} radiative transitions and 41 Mg~{\sc ii} terms 
(up to 11$g$~$^{2}$G at 117639.5~cm$^{-1}$) with 210 Mg~{\sc ii} 
radiative transitions. Regarding the photoionization rates, 
the cross-section data taken from TOPbase (Cunto \& Mendoza 1992) were 
used for the lower 10 Mg~{\sc i} terms and 12 Mg~{\sc ii} terms 
(while hydrogenic approximation was assumed for all other higher terms). 
As to the collisional rates, the recipe described in section~3.1.3 
of Takeda (1991) was followed.

\setcounter{table}{2}
\begin{table}[h]
\begin{minipage}{70mm}
\begin{center}
\caption{Non-LTE calculations done in this study.}
\small
\begin{tabular}{c@{ }cr@{ }rl}
\hline\hline\noalign{\smallskip}
Elem. & $Z^{\#}$ & [X/H]$_{\rm A}^{*}$ & [X/H]$_{\rm B}^{*}$ & References$^{\dagger}$ \\
\hline\noalign{\smallskip}
 He   &  2  &  $-1.0$  & $-0.5$ & Takeda (1994) \\
 C    &  6  &  $-0.8$  & $-0.3$ & Takeda (1992b) \\
 N    &  7  &  $-0.8$  & $-0.3$ & Takeda (1992b) \\
 O    &  8  &  $-0.3$  & $-0.3$ & Takeda (1992a, 2003) \\
 Ne   & 10  &  $-1.0$  & $-0.3$ & Takeda et al. (2010) \\
 Na   & 11  &  $ 0.0$  & $ 0.0$ & Takeda \& Takada-Hidai (1994) \\
 Mg   & 12  &  $ 0.0$  & $ 0.0$ & this paper (Sect.~4.2)\\
 Al   & 13  &  $-1.5$  & $+0.2$ & Takeda (2023) \\
 Si   & 14  &  $+0.2$  & $+0.2$ & Takeda (2022) \\
 P    & 15  &  $+1.0$  & $ 0.0$ & Takeda (2024b) \\
\hline
\end{tabular}
\end{center}
$^{\#}$Atomic number.\\
$^{*}$Abundances (relative to the solar composition)
assigned in non-LTE calculations for each star (in dex), which were 
(iteratively) chosen so that they may be consistent with the final results of 
non-LTE abundances.\\
$^{\dagger}$These papers (and the references quoted therein) may be consulted
for more details about the calculations (e.g., adopted model atoms).
\end{minipage}
\end{table}

\subsection{Abundance results}

After inspection of the list of available lines, it was decided to 
focus on the following 34 species of 28 elements:
He~{\sc i}, C~{\sc i}, N~{\sc i}, O~{\sc i}, Ne~{\sc i},
Na~{\sc i}, Mg~{\sc i}, Mg~{\sc ii}, Al~{\sc i}, Al~{\sc ii},
Si~{\sc ii}, P~{\sc ii}, S~{\sc ii}, Ar~{\sc i}, Ca~{\sc i}, 
Ca~{\sc ii}, Sc~{\sc ii}, Ti~{\sc ii}, V~{\sc ii}, Cr~{\sc i},
Cr~{\sc ii}, Mn~{\sc i}, Mn~{\sc ii}, Fe~{\sc i}, Fe~{\sc ii}, 
Ni~{\sc ii}, Zn~{\sc i}, Sr~{\sc ii}, Y~{\sc ii}, Zr~{\sc ii},
Xe~{\sc ii}, Ba~{\sc ii}, Ce~{\sc ii}, and Nd~{\sc iii}.
By using the equivalent widths of spectral lines or line features
(derived either by direct measurement or synthetic-fitting),
the abundances of these elements were determined by using
Kurucz's (1993) WIDTH9 program, which was considerably modified
by the author (e.g., treatment of merged multi-component features, 
taking into account the non-LTE effect, etc.).
The resulting detailed line-by-line abundances derived for A and B 
(and their line-averaged values), along with the observed 
$W_{\lambda}$ as well as atomic data of spectral lines, are 
presented in ``abundresults.dat'' of online material.

Table~4 presents the mean results averaged over the available lines 
for each species: $\langle$[X/H]$_{\rm A}\rangle$ or
$\langle$[X/H]$_{\rm B}\rangle$ (mean abundance relative to 
the Sun; i.e., line-average of $A_{i} - A_{\odot}$), and 
$\langle \Delta A^{\rm X}_{\rm A-B}\rangle$ (mean of line-by-line
difference between A and B; i.e., line-average of
$A_{i,{\rm A}} - A_{i,{\rm B}}$). The reference solar abundances 
($A_{\odot}$; given in the 3rd column of Table~4)\footnote{It should 
be noted that these data may be somewhat outdated as compared to 
more recent compilations. For example, Asplund et al.'s (2009) solar 
CNO abundances are by $\sim$~0.2--0.3~dex lower than the values adopted here.} 
were taken from Anders and Grevesse (1989) (excepting Fe, for which the value of 
7.50 was used), in order to maintain consistency with the author's 
previous studies. 

In cases where abundances could not be determined because no lines
were measurable, upper-limit abundances were derived by using 
the following lines:
Ne~{\sc i} 6402.248 (1.7), Al~{\sc i} 3944.006 (1.6), Al~{\sc ii} 4663.046 (1.4),
Ar~{\sc i} 8115.311 (2.9), Sc~{\sc ii} 4246.822 (1.5), and Zn~{\sc i} 4722.153 (1.4)
for AR~Aur~A, 
Mn~{\sc i} 4783.430 (1.5), Xe~{\sc ii} 4844.330 (1.5), Ce~{\sc ii} 4711.293 (1.5),
and Nd~{\sc iii} 5127.044 (1.5) for AR~Aur~B.
Here, the values in parentheses are the upper limit equivalent widths
($W^{\rm ul}_{\lambda}$) in m\AA\ estimated by the following relation
\begin{equation}
W^{\rm ul}_{\lambda} = k \; \lambda \;  (v_{\rm w}/c) \left. \middle / \right. {\rm (S/N)},
\end{equation}
where $k (=2)$ is the empirically assigned factor, $\lambda$ is the wavelength,
$v_{\rm w} (\equiv 1.56 \, v_{\rm e}\sin i = 1.56 \times 23) = 36$~km~$^{-1}$ is 
the full-width at half maximum of the rotational broadening function (in velocity unit), 
$c$ is the speed of light, and S/N is the signal-to-noise ratio evaluated by the 
$\lambda$-dependent 3rd-order polynomial depicted in Figure~1c and 1d.  

As given in ``abundresults.dat'', the non-LTE corrections (derived for 
$Z \le 15$ elements)  are considerably different from line to line. 
The mean non-LTE abundance corrections averaged over lines 
($\langle A_{\rm NLTE}-A_{\rm LTE}\rangle$) for A/B are 
$-0.06/-0.07$ (He~{\sc i}), $+0.04/-0.07$ (C~{\sc i}), $-0.21/-0.44$ (N~{\sc i}), 
$-0.49/-0.47$ (O~{\sc i}), $\cdots/-0.07$ (Ne~{\sc i}), $-0.26/-0.32$ (Na~{\sc i}), 
$+0.03/+0.01$ (Mg~{\sc i}), $-0.04/-0.04$ (Mg~{\sc ii}), $\cdots/+0.35$ (Al~{\sc i}),
$\cdots/-0.01$ (Al~{\sc ii}), $-0.03/-0.02$ (Si~{\sc ii}), $-0.10/-0.24$ (P~{\sc ii}).
Regarding the abundances derived for other $Z > 15$ species, for which LTE was
assumed in this study, little can be said about their non-LTE effects. 
We may speculate, however, that the abundances of some elements (e.g., Ca, Sr, Zr,
Ba) may suffer appreciable non-LTE corrections (as much as several tenths dex) by
consulting the recent work of Mashonkina et al. (2020), who studied the non-LTE
effects in deriving the abundances of 14 elements (from He to Nd) for nine 
A- and B-type stars.

The sensitivities of abundance results to changing the atmospheric parameters 
($T_{\rm eff}$, $v_{\rm t}$, and $\log g$) are also presented in Table~4,
where we can see from this table that the impact of $T_{\rm eff}$ is more 
appreciable than the other two. 
That is, abundance errors due to typical uncertainties in $T_{\rm eff}$ 
($\sim \pm 200$~K; cf. Table~2) are $\la$~0.1--0.2~dex, while those due to
$v_{\rm t}$ ($\pm 0.2$~km~s$^{-1}$) as well as $\log g$ ($\pm 0.05$~dex)
are only a few hundredths dex in most cases (note that the $\Delta^{Tvg}$ 
values given in Table~4 correspond to perturbations of $T_{\rm eff}$ by 
250~K, $v_{\rm t}$ by 0.4~km~s$^{-1}$, and $\log g$ by 0.1~dex; i.e., 
twice as large as the typical ambiguities for the latter two).

\setcounter{table}{3}
\begin{table*}[h]
\begin{minipage}{160mm}
\begin{center}
\caption{Results of elemental abundances for AR~Aur A and B.}
\small
\begin{tabular}{ccccc@{ }cc@{ }cc@{ }cc@{ }cc@{ }cc@{ }c
}\hline\hline\noalign{\smallskip}
$Z$ & Species & $A_{\odot}$ & $N$ & $\langle$[X]$_{\rm A}\rangle$  & 
$N_{\rm A}$ & $\langle$[X]$_{\rm B}\rangle$  & $N_{\rm B}$ &  
$\langle \Delta A^{\rm X}_{\rm A-B}\rangle$ & $N_{\rm A-B}$ &
$\Delta^{T-}_{\rm A}$ & $\Delta^{T+}_{\rm B}$ &
$\Delta^{v-}_{\rm A}$ & $\Delta^{v+}_{\rm B}$ &
$\Delta^{g+}_{\rm A}$ & $\Delta^{g+}_{\rm B}$ \\
(1) & (2) & (3) & (4) & (5) & (6) & (7) & (8) & (9) & (10) &
(11) & (12) & (13) & (14) & (15) & (16) \\ 
\hline\noalign{\smallskip}
\multicolumn{16}{c}{(non-LTE analysis)}\\
 2 & He~{\sc i}  & 11.00 &  6 &  $-$1.12&(  6)&  $-$0.55& (  6)&  $-$0.57& (  6)&    +14& $-$15&  +02& $-$01&  +06& +06 \\
 6 & C~{\sc i}   &  8.56 &  7 &  $-$0.78&(  4)&  $-$0.37& (  7)&  $-$0.34& (  4)&    $-$09& +10&  $-$00& +00&  $-$02& $-$02 \\
 7 & N~{\sc i}   &  8.05 & 11 &  $-$1.68&(  3)&  $-$0.54& ( 10)&  $-$1.17& (  3)&    $-$04& +03&  $-$03& +00&  +00& +00 \\
 8 & O~{\sc i}   &  8.93 &  9 &  $-$0.36&(  9)&  $-$0.37& (  9)&  $-$0.03& (  8)&    $-$02& +01&  $-$00& +00&  $-$00& $-$00 \\
10 & Ne~{\sc i}  &  8.09 &  1 &$<-0.8$&(  0)&  $-$0.29& (  1)&$<-0.5$& (  0)&    $\cdots$& $-$13&  $\cdots$& +01&  $\cdots$& +05 \\
11 & Na~{\sc i}  &  6.33 &  4 &  $-$0.08&(  4)&  +0.02& (  4)&  $-$0.11& (  4)&    $-$10& +12&  $-$01& $-$00&  $-$02& $-$03 \\
12 & Mg~{\sc i}  &  7.58 &  7 &  +0.09&(  6)&  +0.15& (  5)&  $-$0.06& (  6)&    $-$14& +14&  +01& $-$01&  $-$04& $-$04 \\
12 & Mg~{\sc ii} &  7.58 &  8 &  $-$0.09&(  8)&  $-$0.02& (  8)&  $-$0.07& (  8)&    +01& $-$00&  +01& $-$01&  +00& $-$00 \\
13 & Al~{\sc i}  &  6.47 &  2 &$<-1.3$&(  0)&  +0.28& (  2)&$<-1.7$& (  0)&    $\cdots$& +13&  $\cdots$& $-$01&  $\cdots$& $-$00 \\
13 & Al~{\sc ii} &  6.47 &  6 &$<-1.5$&(  0)&  +0.16& (  6)&$<-1.7$& (  0)&    $\cdots$& $-$07&  $\cdots$& +01&  $\cdots$& +03 \\
14 & Si~{\sc ii} &  7.55 &  9 &  +0.20&(  9)&  +0.07& (  7)&  +0.07& (  7)&    +04& $-$04&  +02& $-$01&  +03& +03 \\
15 & P~{\sc ii}  &  5.45 &  5 &  +1.26&(  5)&  +0.12& (  1)&  +0.97& (  1)&    +07& $-$07&  $-$00& +00&  +04& +04 \\
\hline\noalign{\smallskip}
\multicolumn{16}{c}{(LTE analysis)}\\
16 & S~{\sc ii}  &  7.21 &  2 &  $-$0.29&(  2)&  $-$0.02& (  2)&  $-$0.27& (  2)&    +10& $-$11&  +00& $-$00&  +06& +06 \\
18 & Ar~{\sc i}  &  6.56 &  2 &$<-0.3$&(  0)&  +0.37& (  2)&$<-0.5$& (  0)&    $\cdots$& +01&  $\cdots$& +01&  $\cdots$& +00 \\
20 & Ca~{\sc i}  &  6.36 &  1 &  $-$0.27&(  1)&  $-$0.22& (  1)&  $-$0.05& (  1)&    $-$24& +25&  +00& $-$00&  $-$07& $-$07 \\
20 & Ca~{\sc ii} &  6.36 &  7 &  $-$0.20&(  6)&  $-$0.32& (  6)&  +0.18& (  6)&    $-$10& +11&  +01& $-$01&  $-$02& $-$02 \\
21 & Sc~{\sc ii} &  3.10 &  1 &$<-1.4$&(  0)&  $-$0.72& (  1)&$<-0.7$& (  0)&    $\cdots$& +14&  $\cdots$& $-$00&  $\cdots$& $-$00 \\
22 & Ti~{\sc ii} &  4.99 & 59 &  +0.69&( 50)&  +0.08& ( 42)&  +0.62& ( 39)&    $-$10& +10&  +04& $-$01&  +01& +01 \\
23 & V~{\sc ii}  &  4.00 &  4 &  +0.25&(  1)&  +0.39& (  4)&  $-$0.09& (  1)&    $-$07& +07&  +01& $-$01&  +02& +02 \\
24 & Cr~{\sc i}  &  5.67 &  3 &  +0.65&(  2)&  +0.43& (  2)&  +0.01& (  1)&    $-$16& +17&  +01& $-$00&  $-$03& $-$03 \\
24 & Cr~{\sc ii} &  5.67 & 47 &  +0.50&( 35)&  +0.39& ( 41)&  +0.12& ( 31)&    $-$04& +04&  +03& $-$01&  +03& +03 \\
25 & Mn~{\sc i}  &  5.39 &  2 &  +1.25&(  2)&$<+0.2$& (  0)&$>+1.1$& (  0)&    $-$15& $\cdots$&  +00& $\cdots$&  $-$03& $\cdots$ \\
25 & Mn~{\sc ii} &  5.39 & 38 &  +1.31&( 33)&  +0.39& (  7)&  +0.95& (  6)&    $-$02& +03&  +02& $-$00&  +03& +03 \\
26 & Fe~{\sc i}  &  7.50 &101 &  +0.56&( 60)&  +0.31& ( 88)&  +0.26& ( 46)&    $-$13& +14&  +02& $-$01&  $-$03& $-$03 \\
26 & Fe~{\sc ii} &  7.50 &339 &  +0.62&(304)&  +0.30& (240)&  +0.31& (211)&    +01& $-$01&  +03& $-$01&  +03& +03 \\
28 & Ni~{\sc ii} &  6.25 &  7 &  $-$0.79&(  1)&  +0.62& (  7)&  $-$1.39& (  1)&    $-$01& $-$01&  +01& $-$01&  +04& +03 \\
30 & Zn~{\sc i}  &  4.60 &  1 &$<+0.5$&(  0)&  +0.78& (  1)&$<-0.3$& (  0)&    $\cdots$& +12&  $\cdots$& $-$00&  $\cdots$& $-$03 \\
38 & Sr~{\sc ii} &  2.90 &  2 &  +1.81&(  2)&  +0.89& (  2)&  +0.92& (  2)&    $-$17& +18&  +23& $-$11&  $-$02& $-$03 \\
39 & Y~{\sc ii}  &  2.24 & 11 &  +2.37&( 11)&  +0.75& (  1)&  +1.45& (  1)&    $-$16& +16&  +09& $-$00&  $-$01& $-$01 \\
40 & Zr~{\sc ii} &  2.60 &  9 &  +1.67&(  7)&  +0.77& (  3)&  +0.91& (  3)&    $-$13& +13&  +04& $-$01&  +00& +01 \\
54 & Xe~{\sc ii} &  2.23 &  1 &  +4.70&(  1)&$<+3.9$& (  0)&$>+0.8$& (  0)&    +11& $\cdots$&  +04& $\cdots$&  +09& $\cdots$ \\
56 & Ba~{\sc ii} &  2.13 &  3 &  +1.03&(  2)&  +1.56& (  2)&  $-$0.65& (  1)&    $-$16& +18&  +02& $-$07&  $-$02& $-$03 \\
58 & Ce~{\sc ii} &  1.55 &  1 &  +5.01&(  1)&$<+3.9$& (  0)&$>+1.1$& (  0)&    $-$14& $\cdots$&  +01& $\cdots$&  $-$02& $\cdots$ \\
60 & Nd~{\sc iii}&  1.50 & 12 &  +2.51&( 10)&$<+0.7$& (  0)&$>+1.9$& (  0)&    $-$04& $\cdots$&  +04& $\cdots$&  +04& $\cdots$ \\
\hline
\end{tabular}
\end{center}
(1) Atomic number. (2) Element species. (3) Reference solar abundances (in the 
usual normalization of H = 12.00), which are taken from Anders \& Grevesse's (1989) 
compilation (except for Fe, for which 7.50 is adopted). 
(4) Number of lines adopted for this species.
(5) Mean of [X/H]$_{\rm A}$ (relative abundance for A in comparison with the Sun) 
averaged over lines. (6) Actual number of lines used for deriving $\langle$[X/H]$_{\rm A}\rangle$.
(7) Mean of [X/H]$_{\rm B}$. (8) Actual number of lines employed for $\langle$[X/H]$_{\rm B}\rangle$.
(9) Mean of $\Delta A^{\rm X}_{{\rm A}-{\rm B}}$ (differential line-by-line abundance between
A and B) averaged over lines. (10) Actual number of lines used for calculating
$\langle \Delta A^{\rm X}_{\rm A-B}\rangle$. 
(11) Abundance change for A in response to $\Delta T_{\rm eff} = -250$~K.  
(12) Abundance change for B in response to $\Delta T_{\rm eff} = +250$~K.   
(13) Abundance change for A in response to $\Delta v_{\rm t} = -0.4$~km~s$^{-1}$.  
(14) Abundance change for B in response to $\Delta v_{\rm t} = +0.4$~km~s$^{-1}$.
(15) Abundance change for A in response to $\Delta \log g = +0.1$~dex.  
(16) Abundance change for B in response to $\Delta \log g = +0.1$~dex.\\
All abundance-related data ($A_{\odot}$, $\langle$[X/H]$\rangle$, $\Delta$) are 
in unit of dex. See Section~4.3 regarding how the upper limit abundance was estimated
for unmeasurable cases. Note that only the 1st and 2nd decimals are shown 
in the data of (11)--(16) (i.e., they should be divided by 100). 
\end{minipage}
\end{table*}

\section{Discussion}

\subsection{$T_{\rm eff}$ and $v_{\rm t}$ for A and B}

The values of $T_{\rm eff}$ and $v_{\rm t}$ were established by 
using Fe~{\sc ii} lines in this study (Sect.~3). Comparing these 
spectroscopic $T_{\rm eff,A}/T_{\rm eff,B}$ (11150/10650~K) with 
Nordstr\"{o}m \& Johansen's (1994) photometric SED-based determinations 
(10950/10350~K), we can see that, while our results tend to be somewhat 
higher than theirs by $\sim$~200--300~K in the absolute scale, the 
relative differences between A and B ($\Delta T_{\rm eff,A-B}$ = +500~K 
in this study and +600~K in their paper) are reasonably consistent 
with each other. 
   
Regarding $v_{\rm t}$, for which determinations for this AR~Aur system 
have failed so far, the results (0.9~km~s$^{-1}$ for A and 0.1~km~s$^{-1}$ 
for B) concluded by the analysis of Fe~{\sc ii} lines in Section~3 
may be regarded as significant. It should be noted that similar inequality 
relation of $v_{\rm t,A} > v_{\rm t,B}$ (i.e., larger $v_{\rm t}$ for 
higher $T_{\rm eff}$) was obtained also 
from Fe~{\sc i} lines (though not adopted; cf. Table~1).
Therefore, this increase of $v_{\rm t}$ with $T_{\rm eff}$ in AR~Aur A and B 
(late B-type stars) does not follow the trend seen in early A-type stars 
where $v_{\rm t}$ tends to progressively decrease with $T_{\rm eff}$ 
(see, e.g, Takeda et al. 2008). 
This characteristics of appreciably smaller $v_{\rm t}$ values ($\sim$~0--1~km~s$^{-1}$) 
in late-B dwarfs (in comparison with early-A dwarfs which have typical $v_{\rm t}$ 
of $\sim$~2~km~s$^{-1}$) is consistent with the results of several spectroscopic 
studies done by other investigators (e.g., Allen 1998; Hubrig et al. 1999; 
Saffe et al. 2011).\footnote{
The literature $v_{\rm t}$ values of late-B type stars compiled by Sadakane 
(1990; cf. table~4 therein) are as large as $\sim$~1--2~km~s$^{-1}$ with 
a roughly decreasing tendency with $T_{\rm eff}$ (similarly to the case of 
early A-type stars), which may appear to contradict the trend mentioned here. 
Note, however, since class V (dwarfs), class IV (subgiants), and class III 
(giants) are mixed in Sadakane's (1990) sample, any definite conclusion can 
not be made from such an inhomogeneous data set.} 
Considering that A and B are quite similar
in terms of other parameters, this fact of discrepant $v_{\rm t}$ under only 
a small difference in $T_{\rm eff}$ might be a clue to understand the 
nature of microturbulence in late B-type stars.

\subsection{Trends of chemical abundances}

Based on the results in Table~4, 
$\langle$[X/H]$_{\rm A}\rangle$, $\langle$[X/H]$_{\rm B}\rangle$, and 
$\langle \Delta A^{\rm X}_{\rm A-B}\rangle$ are plotted against $Z$ in 
Figure~7a, 7b, and 7c. Considering the impact of uncertainties in atmospheric 
parameters (cf. Table~2 and Table~4) and the size of standard deviations in 
the averages (cf. ``abundresults.dat''), typical statistical errors involved 
with the data symbols in Figure~7 may be estimated as $\pm \la$~0.1--0.2~dex. 
It should be kept mind, however, that additional systematic errors might be 
possible in [X/H] values related to other factors (e.g., reference solar 
abundances, $gf$ values, non-LTE effect for $Z>15$ elements, etc.),
while $\Delta A^{\rm X}_{\rm A-B}$ values (line-by-line differential
abundances) are distinctly more advantageous because they are almost 
irrelevant to such systematic error sources.

Several notable characteristics are observed by inspecting Figure~7,
as summarized below (the symbols ``$\langle$'' and  ``$\rangle$'' to indicate 
average values are omitted for simplicity):
\begin{itemize}
\item
Roughly speaking, [X/H] (departure from the solar composition) 
tends to increase with $Z$ in the global 
sense for both A and B (Fig.~7a and 7b); i.e., [X/H]~$<0$ at $Z \la 10$, 
[X/H]~$\sim 0$ at $10 \la Z \la 20$, and [X/H]~$>0$ at $Z \ga 20$.
Quantitatively, the degree of $Z$-dependence (or the slope of
linear-regression line) is steeper for A than for B. As a result,
$\Delta A^{\rm X}_{\rm A-B}$ also shows a similar $Z$-dependent trend (Fig.~7c)
\item
Yet, the dispersion of [X/H]$_{\rm A}$ is considerably large, because some 
elements show conspicuous deviations from the global trend by as much as
$\sim \pm 1$~dex; that is, the pronounced deficits of N, Al, Sc, and Ni;
or the prominent excesses of P and Mn (and rare earths). Since these are 
the characteristics of chemically peculiar stars of HgMn type (e.g., 
Ghazaryan \& Alecian 2016), we can state that AR~Aur~A is surely a HgMn star.
\item
In contrast, the $Z$-dependence of [X/H]$_{\rm B}$ is apparently more tight,
which almost linearly correlate with $Z$ ([X/H]$_{\rm B} \simeq -0.6 + 0.04 Z$)
as depicted in Figure~7b. Actually, it is only Sc and high-$Z$ elements 
(Xe, Ce, Nd) that markedly deviate from this relation.
Folsom et al. (2010) suggested AR~Aur B to belong to a weak Am star. 
Admittedly, the ``global'' trend of its abundance pattern is
similar to that shown by Am-Fm stars (see, e.g., fig.~5 in Smith 1996).
However, on a close inspection, the abundance peculiarities of Am stars 
do not exhibit such a remarkable $Z$-dependent near-linearity. 
For example, [X/H] values of Sirius (well-known hot Am star) show 
appreciable local fluctuations (cf. fig.~4 in Michaud et al. 2011), 
though surely tending to increase with $Z$ in the global sense.
Therefore, a new type specific to late B-type stars (Bm stars?) might 
as well be assigned to this kind of peculiarity in AR~Aur~B.
\end{itemize} 

\begin{figure}[h]
\begin{minipage}{70mm}
\begin{center}
  \includegraphics[width=7.0cm]{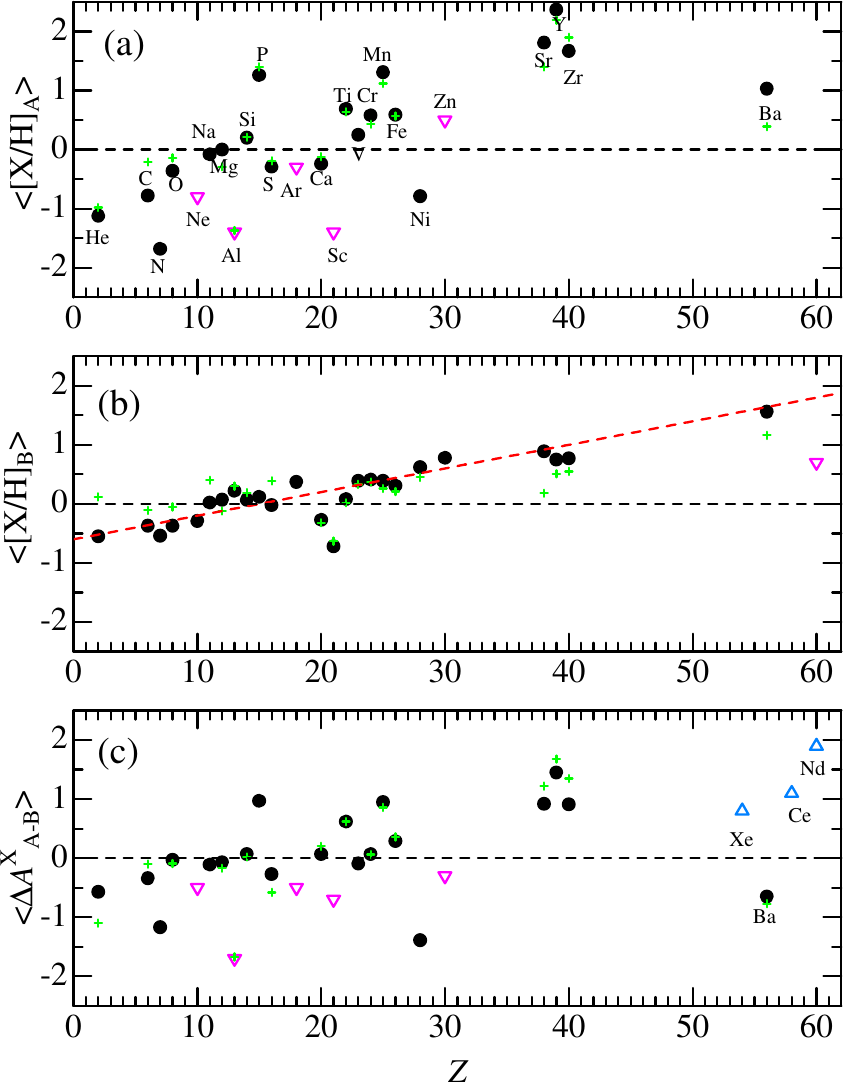}
\end{center}
   \caption{
Plotted against $Z$ (atomic number) in black filled bullets are  
(a) $\langle$[X/H]$_{\rm A}\rangle$ (averaged 
differential abundance of element X relative to the Sun for AR~Aur~A), 
(b) $\langle$[X/H]$_{\rm B}\rangle$  (ditto for AR~Aur~B), and 
(c) $\langle \Delta A^{\rm X}_{{\rm A}-{\rm B}}\rangle$ (averaged 
line-by-line differential abundance of element X between A and B), 
for 28 elements (He, C, N, O, Ne, Na, Mg, Al, Si, P, 
S, Ar, Ca, Sc, Ti, V, Cr, Mn, Fe, Ni, Zn, Sr, Y, Zr, Xe, Ba, Ce, and Nd) 
based on the data in Table~4. Regarding the elements where two results
of different ionization stages are available (Mg, Al, Ca, Mn, Fe), 
their mean values are adopted here (exceptionally, the result for 
Mn~{\sc ii} is used for Mn because that for Mn~{\sc i} is only the upper limit).
The upper limit and lower limit values are indicated by pink open 
inverse triangles and blue open normal triangles, respectively. 
In panel (b), the linear relation, 
$\langle$[X/H]$_{\rm B}\rangle = -0.6 + 0.04 Z$, is depicted by a dashed line.
Note that in panels (a) and (b), some results for high-$Z$ elements 
(Xe, Ce, Nd) are outside of the plot range.
For reference, Folsom et al.'s (2010) results of [X/H]$_{\rm A}$,
[X/H]$_{\rm B}$, and [X/H]$_{\rm A}-$[X/H]$_{\rm B}$ are also shown
by light-green small crosses in each of the panels.  
} 
   \label{Fig7}
\end{minipage}
\end{figure}

\subsection{Comparison with previous results}

Let us compare the abundance characteristics of AR~Aur A and B resulting 
from this investigation (Sect.~5.2)  with those reported by previous 
chemical abundance studies for this system (Sect.~1).

Khokhlova et al.'s (1995) work was the first attempt to determine the 
abundances of A and B for various elements. Although high precision
would not have been expected because of methodological disadvantages 
in comparison with the present-day standard (i.e., use of photographic 
spectrogram, difficulty in measuring $W_{\lambda}$ on double-lined spectra, 
classical curve-of-growth analysis), their conclusions are qualitatively 
consistent with our results (typical HgMn-type peculiarity in A, less 
pronounced anomalies of different kind in B). The same applies also to 
Ryabchikova's (1998) results of reanalyzing Khokhlova et al.'s $W_{\lambda}$ 
data using model atmospheres. Folsom et al. (2010; sect.~5 therein) stated 
that the abundances obtained by them are in reasonable agreement with 
those of Ryabchikova (1998), except that their results are appreciably 
lower than the latter in some specific elements (C, Mn, Sr, Y, and Pt 
for A; Ti, Mn, Fe, and Ba for B).

The elemental abundances of AR~Aur A and B determined by Folsom et al. 
(2010) based on the spectrum synthesis technique are more extensive  
and probably more reliable than any other previous determinations.
Therefore, their [X/H]$_{\rm A}$ and [X/H]$_{\rm B}$ values (taken from 
table~3 therein) are overplotted in Figure~7 for comparison. 
As seen from this figure, their results (light-green crosses) are 
satisfactorily consistent with those of this study (black bullets), 
though some minor discrepancies are observed in several elements 
(such as He, C, O, Sr, Y. Zr, and Ba). This may be attributed to several 
factors (e.g., non-LTE effect, different choice of fiducial solar 
abundance, difference in the adopted microturbulence). 
 
Although the main purpose of Paper~I was to establish the abundances of 
C, N, and O for both of the binary components, those of Na, Si, Ca, 
Ti, Fe, and Ba (obtained as by-products) were also presented. 
Those [X/H] values (cf. table~7 therein) determined for AR~Aur A and B 
do not necessarily agree well with the present results. This is because 
(i) there were some cases where the lines used in Paper~I (primary attention 
being paid to A-type stars) were too weak for late B-type stars, and
(ii) microturbulences ($v_{\rm t,A}, v_{\rm t,B})$ of (1.0, 
1.6~km~s$^{-1}$) employed in Paper~I were different from the values (0.9, 
0.1~km~s$^{-1}$) adopted in this paper (especially for B).
As such, considerable discrepancies (larger than 0.3~dex) are seen in
[C/H]$_{\rm A}$, [C/H]$_{\rm B}$, [Na/H]$_{\rm A}$, [Si/H]$_{\rm B}$,
[Ca/H]$_{\rm A}$, [Ba/H]$_{\rm A}$, and [Ba/H]$_{\rm B}$.
In any case, the abundances of AR~Aur derived in Paper~I should be 
replaced by the new results of this investigation presented in Table~4.

\subsection{Origin of compositional differences between A and B}

Now that the nature of surface abundances for AR~Aur A and B has been
elucidated, the key issue to be considered is ``what is the cause of 
such appreciable chemical differences between these two similar stars?''

Whereas the chemical abberations of component B ($T_{\rm eff} = 10650$~K) 
are rather weak and organized ([X/H] is almost linearly dependent upon $Z$), 
the slightly hotter A ($T_{\rm eff} = 11150$~K) in turn exhibits wildly 
conspicuous HgMn-type abundance anomalies (strong deficit of N, Al, Sc, Ni; 
prominent excess of P, Mn). Which physical process is responsible for 
this remarkable transmutation of surface chemistry for only a small 
($\sim 500$~K) change in $T_{\rm eff}$?

If this transition is caused by the difference of $T_{\rm eff}$, 
HgMn peculiarity must be a very $T_{\rm eff}$-sensitive phenomenon
in the sense that it is suddenly triggered once $T_{\rm eff}$
exceeds a critical value at $\simeq 11000$~K.
Admittedly, the existence of such a $T_{\rm eff}$ limit is consistent 
with the statistical study of late B-type stars by Wolff \& Preston (1978), 
who found that HgMn stars are observed only at 
11000~K~$\la T_{\rm eff} \la $~16000~K. 
Yet, it may still be wondered which kind of mechanism is underlying 
in such abrupt surface abundance changes at this critical $T_{\rm eff}$.

We should recall that $v_{\rm t}$ is appreciably discordant between 
A (0.9~km~s$^{-1}$) and B (0.1~km~s$^{-1}$). One may question whether 
this difference has something to do with the distinction of abundance 
characteristics between A and B. However, if more stable atmosphere is 
favored for the emergence of chemical peculiarity (as often postulated 
in element diffusion theory), this inequality in $v_{\rm t}$ (i.e., 
stronger anomaly for larger turbulence) may be contradictory to 
our intuitive picture.   

Another issue worth consideration is the evolutionary status of the AR~Aur 
system. Nordstr\"{o}m \& Johansen (1994) concluded from the unusual 
inequality of radii ($R_{\rm B} > R_{\rm A}$) despite of the mass 
difference ($M_{\rm B} < M_{\rm A}$) that B is likely to be still in the 
pre-main sequence stage of contraction phase, while A is already on the main 
sequence of hydrogen burning phase. However, their conclusion is unconvincing 
and questionable, as separately discussed in Appendix~A,
But, if this scenario is really the case, such a difference in the evolutionary 
phase may affect the surface chemistry. For example, strong abundance anomaly 
would take place quickly after a star has arrived at the main sequence 
where atmospheric stability is realized (corresponding to A), while conspicuous 
chemical peculiarity would be difficult to develop during the unstable contraction 
phase (corresponding to B).

In any event, the abundance characteristics of AR~Aur A and B 
(especially their marked differences from each other) established 
in this investigation may serve as important observational facts 
for any theory trying to explain the chemical anomaly of late B-type stars.  

\section{Summary and conclusion}

The close binary system AR~Aur A+B is of astrophysical importance, because surface 
compositions of both are known to exhibit appreciable differences despite that 
they are similar B9V and B9.5V main-sequence stars.
By investigating the chemical abundances characteristics of A and B along with 
the differences of stellar atmospheric parameters, we may gain useful 
physical insight into the mechanism of how and which chemical anomaly takes place 
in late B-type stars.

Although stellar parameters of this system are comparatively well established
by making use of the merit of spectroscopic/eclipsing binary, reliable 
spectroscopic determinations of $T_{\rm eff}$ and $v_{\rm t}$ (important 
atmospheric parameters for spectroscopic analysis) have not yet been done. 
Likewise, publications of extensive elemental abundance study for 
both A and B based on high-quality data are still insufficient, reflecting 
the difficulty of analyzing complex double-lined spectra of AR~Aur.  

Motivated by this situation, the author decided to carry out a detailed 
spectroscopic study for each component, in order to determine the key 
atmospheric parameters and elemental abundances of A and B as precisely 
as possible and to examine how they compare with each other.

Regarding the basic observational material, the spectrum-disentangle technique
was applied to a set of original double-line spectra taken at different orbital 
phases to obtain the decomposed spectra of A and B. Based on these disentangled 
spectra (covering 3900--9200~\AA), many lines judged to be usable were identified 
and their $W_{\lambda}$ were measured by the direct function-fitting. 
In addition, $W_{\lambda}$ values of important line features (even if they consist
of complex multi-components) were evaluated by applying the spectrum-synthesis 
technique.
 
The values of ($T_{\rm eff}$, $v_{\rm t}$) were determined from Fe~{\sc ii} 
lines by requiring that abundances do not show any systematic dependence upon
$W_{\lambda}$ and $\chi_{\rm low}$, which turned out (11150~K, 0.9~km~s$^{-1}$) 
and (10650~K, 0.1~km~s$^{-1}$) for A and B, respectively. 

The chemical abundances of 28 elements (34 species) were derived from 
the $W_{\lambda}$ values of many lines, where the non-LTE effect was taken 
into account for comparatively lighter elements of $Z \le 15$.
The following characteristics were found in the resulting [X/H]$_{\rm A}$ 
and [X/H]$_{\rm B}$. \\
--(1) Qualitatively, a rough $Z$-dependent tendency holds for both A and B that 
light elements ($Z \la 10$ such as He, C, N, O) are underabundant, heavier elements 
($Z \ga 20$; such as Fe group, s-process, rare earths) overabundant, 
and nearly solar for intermediate cases ($10 \la Z \la 20$). 
Likewise, a similar trend of increasing with $Z$ is roughly seen in the 
differential abundances between A and B ($\Delta A^{\rm X}_{{\rm A}-{\rm B}}$), 
since the peculiarity is quantitatively more conspicuous in A than in B.\\
--(2) Yet, regarding the hotter A, several elements show strikingly 
large peculiarities (e.g., very deficient N, Al, Sc, Ni or very 
overabundant P, Mn; leading to a considerable dispersion). 
These are the well-known characteristics of HgMn stars.\\
--(3) In contrast, the situation in the cooler B is apparently more simple 
in the sense that the progressive $Z$-dependence of [X/H]$_{\rm B}$ 
almost follows the linear relation ([X/H]$_{\rm B} = -0.6 + 0.04 Z$) 
with only a small dispersion. Therefore, B shows a comparatively weak and 
rather organized peculiarity.

Therefore, the next important task would be to clarify the cause why such 
remarkably dissimilar type of chemical peculiarities are observed in these 
two similar late B-type stars (with a small $T_{\rm eff}$ difference 
of $\sim$~500~K), for which further contributions of theoreticians 
are awaited.

\normalem
\begin{acknowledgements}
Data analysis (numerical calculation of spectrum disentangling) was carried 
out on the Multi-wavelength Data Analysis System operated by the Astronomy 
Data Center (ADC), National Astronomical Observatory of Japan.
This investigation has made use the VALD database operated at Uppsala University,
the Institute of Astronomy RAS in Moskow, and the University of Vienna.
\end{acknowledgements}

\section*{Online materials}

This article accompanies the following online materials 
(electronic data files).
See ``ReadMe'' for the details about their contents.
\begin{itemize}
\item
ReadMe 
\item
spectra\_A.txt 
\item
spectra\_B.txt
\item
identlist\_A.dat 
\item
identlist\_B.dat 
\item
abundresults.dat 
\end{itemize}

\clearpage
\appendix

\section{Is AR~Aur~B in pre-main sequence stage?}

Nordstr\"{o}m \& Johansen (1994) concluded that AR~Aur~B is still in the 
contracting phase of pre-main-sequence, whereas AR~Aur~A is already in the 
H-burning phase on the main sequence.
This conclusion is based on the radius ratio they derived for this 
system $R_{\rm B}/R_{\rm A} = 1.02 \pm 0.015 (>1)$.
That is, since the mass ratio is robustly determined from the spectroscopic 
orbital elements as $M_{\rm B}/M_{\rm A} = 2.29/2.48 = 0.925 (<1)$, 
such an inequality relation ($R_{\rm B}/ R_{\rm A} > 1$) is impossible 
if both A and B are main-sequence stars (where $R$ increases with $M$).
Thus, the only solution simultaneously satisfying both conditions
in terms of $R$ and $M$ is to regard that B has not yet reached the 
main sequence but A is already on it. 

However, their argument is not convincing. Since precisely determining 
$R_{\rm B}/R_{\rm A}$ from light curve analysis is difficult for the 
case of AR~Aur (partially eclipsing system with nearly similar components), 
they invoked the observed equivalent width ratio ($W_{\rm B}/W_{\rm A}$) 
of the strong Mg~{\sc ii} 4481 line by making use of the relation
\begin{equation}
\begin{split}
W^{\rm obs}_{\rm B}/W^{\rm obs}_{\rm A} 
& = (W^{\rm int}_{\rm B}/W^{\rm int}_{\rm A})(L_{\rm B}/L_{\rm A}) \\
& = (W^{\rm int}_{\rm B}/W^{\rm int}_{\rm A})(J_{\rm B}/J_{\rm A})(R_{\rm B}/R_{\rm A})^{2}, 
\end{split}
\end{equation}
where $W^{\rm obs}$ and $W^{\rm int}$ are the observed equivalent width
on the double-line spectrum and the intrinsic equivalent width 
($W^{\rm obs} < W^{\rm int}$ because of the dilution effect), 
while $J$ and $L$ are the surface brightness and the luminosity 
at the waveband of the line.
That is, since $W^{\rm obs}_{\rm B}/W^{\rm obs}_{\rm A}$ is directly measurable by 
spectroscopic observations and $J_{\rm B}/J_{\rm A}$ is derived from photometric
solutions, $R_{\rm B}/R_{\rm A}$ can be determined from Equation~(A.1) if 
$W^{\rm int}_{\rm B}/W^{\rm int}_{\rm A}$ is somehow known.

The serious weak point of this method is the necessity of ``assuming'' some
appropriate value for $W^{\rm int}_{\rm A}/W^{\rm int}_{\rm B}$, because
it is unknowable in advance. Nordstr\"{o}m \& Johansen (1994) adopted 
$W^{\rm int}_{\rm A}/W^{\rm int}_{\rm B} = 0.925 \pm 0.005$
(or $W^{\rm int}_{\rm B}/W^{\rm int}_{\rm A} = 1.081 \pm 0.006$)
based on the rough linear relation between $W^{\rm int}_{4481}$ and 
the surface flux (dependent upon the spectral type or $T_{\rm eff}$)
constructed from the data of several late B-type stars (cf. their fig.~10).
In other words, they considered only the difference of $T_{\rm eff}$
between two components, but assumed other parameters affecting 
the strength of Mg~{\sc ii} 4481 line to be the same. This is unjustifiable
because a slight difference in Mg abundances and a large discordance in 
$v_{\rm t}$ actually exist between A and B (as clarified in this study).
As such, their assumed $W^{\rm int}_{\rm B}/W^{\rm int}_{\rm A}$ ratio must 
suffer appreciable systematic errors due to other factors neglected by them 
(declared uncertainty of only 0.5\% is too optimistic).
Accordingly, their final result of $R_{\rm B}/R_{\rm A} = 1.02 \pm 0.015$ 
is not trustworthy, because actual error would be more significant  
(this error of 1.5\% is nothing but due to the uncertainty involved 
in their $W^{\rm obs}_{\rm B}/W^{\rm obs}_{\rm A}$).

It should also be pointed out that, although both components being in 
different evolutionary stages (A: main sequence, B: pre-main sequence) 
may be conceptually possible, it must be a rare incidence (even in 
such a circumstance) to observe them as almost similar stars 
($M_{\rm A} \simeq M_{\rm B}$ and $R_{\rm A} \simeq R_{\rm B}$) 
because evolutionary time scales are markedly different.
Figure~A.1 illustrates how the physical parameters 
($T_{\rm eff}$, $L$, $R$, and $\log g$) of 2.1, 2.2, 2.3, 2.4, 2.5, 2.6, 2.7, 
and 2.8~$M_{\odot}$ stars vary in the pre-main-sequence phase ($ t < 0$) 
and in the main-sequence phase ($t \ge 0$), which are taken from 
Bressan et al.'s (2012) theoretical stellar evolution calculations.
It can be easily seen from this figure that the time scale of parameter 
variations due to evolution is much shorter at $t < 0$ than that at $t \ge 0$.
Admittedly, if the timing of star formation for A and B is slightly different
and the condition of $t_{\rm B} < 0 < t_{\rm A}$ is appropriately satisfied. 
it may be possible to observe $R_{\rm B} > R_{\rm A}$ simultaneously with 
$M_{\rm B} < M_{\rm A}$ (cf. Figure~A.1c). Yet, in order to realize
the near-similarity of $R$ in this case, $t_{\rm A}$ and (especially) 
$t_{\rm B}$ have to be at the pinpointed timing, which would rarely 
happen by coincidence. In short, AR~Aur A and B are too similar to 
regard B as being still in the pre-main-sequence phase.

Although the argument of Nordstr\"{o}m \& Johansen (1994) seems to be 
questionable as described above, it is premature to conclude that it is 
incorrect. This possibility should still be further investigated. 
We should note that Folsom et al. (2010) independently carried out 
a similar analysis using the Mg~{\sc ii} 4481 line and obtained  
the radius ratio $R_{\rm B}/R_{\rm A} = 1.033 \pm 0.005$ (confirming 
the inequality $R_{\rm B}/R_{\rm A} >1$ with even higher precision) 
which is in support of Nordstr\"{o}m \& Johansen's (1994) consequence.  
Unfortunately, since any detailed account is not given 
regarding how they derived this value (i.e., adopted  
$W^{\rm obs}_{\rm B}/W^{\rm obs}_{\rm A}$ or 
$W^{\rm int}_{\rm B}/W^{\rm int}_{\rm A}$, evaluation of errors, etc.), 
it is hardly possible to comment on their result.

In any case, given that this is a difficult problem demanding to detect  
only a slight difference of $R$ with high precision, it is not sufficient to 
invoke only one Mg~{\sc ii} 4481 line to draw any definite conclusion. 
Such an analysis should be done for at least several lines, in order 
to confirm whether similar results are derived from different lines. 

\begin{figure}[h]
\begin{minipage}{70mm}
\begin{center}
  \includegraphics[width=7.0cm]{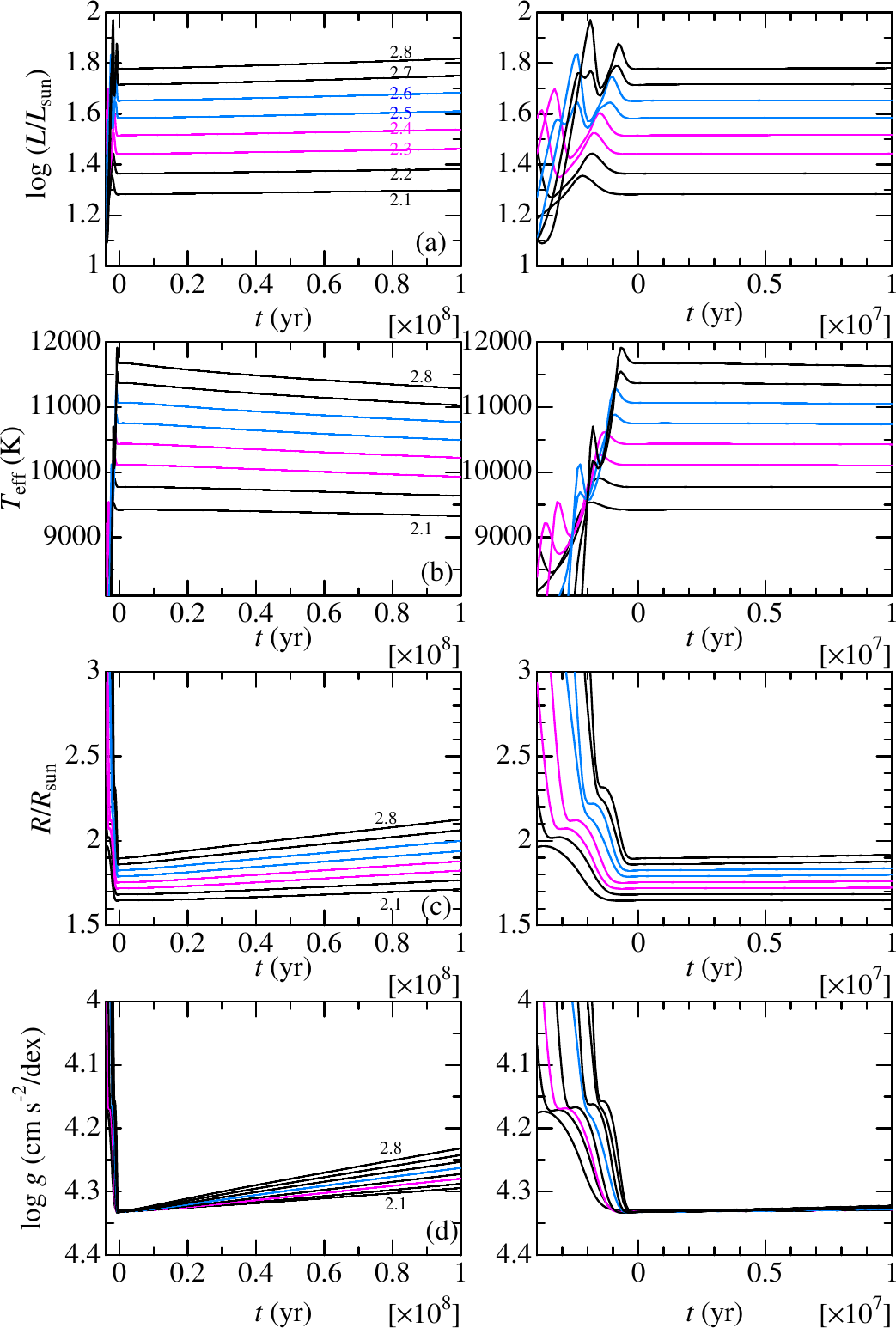}
\end{center}
   \caption{Evolution of physical parameters of main-sequence stars 
(metallicity $Z = 0.02$) with the masses of 2.1, 2.2, 2.3, 2.4, 2.5, 2.6, 
2.7, and 2.8~$M_{\odot}$, based on Bressan et al.'s (2012) PARSEC database.
Plotted against the elapsed time ($t$) are (a) $L$ (bolometric luminosity), 
(b) $T_{\rm eff}$ (effective temperature), (c) $R$ (radius), and 
(d) $\log g$ (surface gravity). Note that the origin ($t=0$) corresponds
to the beginning of H burning in the core  (Zero-Age Main Sequence).
The left and right panels are paired and different only in the covered
time span; the former (global view) is from $-4\times 10^{6}$~yr
to $+1\times 10^{8}$~yr, while the latter (magnified view) is from 
$-4\times 10^{6}$~yr to $+1\times 10^{7}$~yr. 
The results for the mass values almost corresponding to AR~Aur A and B 
(2.5--2.6~$M_{\odot}$ for A, 2.3--2.4~$M_{\odot}$ for B) are colored
in blue and pink, respectively.
Since the results for 2.5 and 2.7~$M_{\odot}$ are not available in the 
original data, those of 2.4/2.6~$M_{\odot}$ and 2.6/2.8~$M_{\odot}$ 
were simply averaged, respectively.
} 
   \label{FigA1}
\end{minipage}
\end{figure}

\end{document}